\documentclass[useAMS,usenatbib,usegraphicx]{mn2e}

\usepackage{color}
\title[Accretion mode of ULX M82 X-2]{Accretion mode of the 
Ultra-Luminous X-ray source \hbox{M82 X-2}}
\author[S. Karino and J. C. Miller]
{S. Karino$^{1}$\thanks{E-mail: karino@ip.kyusan-u.ac.jp} and 
J. C. Miller$^{2}$\thanks{E-mail: john.miller@physics.ox.ac.uk} \\
$^{1}$Faculty of Engineering, Kyushu Sangyo University, 2-3-1 Matsukadai, 
Fukuoka 813-8503, Japan \\
$^{2}$Department of Physics (Astrophysics), University of Oxford, Keble 
Road, Oxford OX1 3RH, UK}

\begin{document}

\date{2016 May 7}

%\pagerange{\pageref{firstpage}--\pageref{lastpage}} \pubyear{2002}

\maketitle

\begin{abstract}

Periodic pulsations have been found in emission from the ultra-luminous 
X-ray source (ULX) M82 X-2, strongly suggesting that the emitter is a 
rotating neutron star rather than a black hole. However, the radiation 
mechanisms and accretion mode involved have not yet been clearly 
established. In this paper, we examine the applicability to this object 
of standard accretion modes for high mass X-ray binaries (HMXBs). We find 
that spherical wind accretion, which drives OB-type HMXBs, cannot apply 
here but that there is a natural explanation in terms of an extension of 
the picture for standard Be-type HMXBs. We show that a neutron star with 
a moderately strong magnetic field, accreting from a disc-shaped wind 
emitted by a Be-companion, could be compatible with the observed relation 
between spin and orbital period. A Roche lobe overflow picture is also 
possible under certain conditions.

\end{abstract}


\section{Introduction}

Ultra-luminous X-ray sources (ULXs) are very bright extragalactic X-ray 
point sources, with observed fluxes which would correspond to 
luminosities greater than $10^{39}$ ergs per second if they were 
radiating isotropically. Since this is above the Eddington limit for 
normal stellar-mass compact objects, it has been widely thought that they 
are associated with intermediate-mass black holes (IMBHs) \citep{F09}, 
although non-spherical accretion and beamed emission could give rise to 
inferred luminosities significantly above the Eddington limit \citep[see, 
for example][]{OM11}, allowing also for lower masses.

\citet{B14} have reported NuSTAR observations of ULX M82 X-2 (also known 
as NuSTAR J095551+6940.8) which reveal periodic changes in the hard X-ray 
luminosity of this source, indicative of a rotating magnetized neutron 
star being involved rather than a black hole. The measured peak flux (in 
the $3-30 \rm{keV}$ band) would correspond to $L_X = 3.7 \times 10^{40} 
\, \rm{erg \, s}^{-1}$ if the radiation were isotropic, and is 
challenging to explain with a neutron star. The period (taken to be the 
neutron-star spin period) was found to be $P_{\rm{s}} = 1.37 \, \rm{s}$, 
with a 2.53-day sinusoidal modulation, interpreted as being an orbital 
period $P_{\rm{orb}}$ corresponding to motion around an unseen companion 
which would be the mass donor in the accreting system. The time 
derivative of the spin period $\dot{P}_{\rm{s}}$ was also measured. 
Values for this coming from different individual observations show 
considerable variations but a relevant underlying spin-up tendency was 
found, with $\dot{P}_{\rm{s}} = -2 \times 10^{-10} \, \rm{s \, s}^{-1}$. 
The mass donor is indicated as having a mass larger than $5.2 M_{\odot}$, 
so that the system should be categorized as a high mass X-ray binary 
(HMXB). Taking canonical neutron star parameters as a rough guide 
($M_{\rm{NS}} = 1.4 M_{\odot}$ and $R_{\rm{NS}} = 10 \, \rm{km}$), the 
luminosity relation $L_{\rm{X}} = G M_{\rm{NS}} \dot{M} 
R_{\rm{NS}}^{-1}$, gives the mass accretion rate corresponding to 
$L_{\rm{X}} = 3.7 \times 10^{40} \rm{erg \, s}^{-1}$ as being $ \dot{M} = 
2.0 \times 10^{20} \rm{g \, s}^{-1} = 3.1 \times 10^{-6} M_{\odot} 
\rm{yr}^{-1} $.

There are three main mechanisms by which the mass transfer might occur: 
(i) via a spherical wind (as for O-type HMXBs), (ii) via a disc-shaped 
wind (as for Be-type HMXBs), or (iii) by Roche lobe overflow (RLOF). 
Because of the large inferred $\dot{M}$, the third option was suggested 
as the mechanism by \citet{B14} and subsequent studies \citep{E14,DPS14}.

Here, we investigate each of these scenarios in turn to see which may be 
appropriate for ULX M82 X-2. In Section 2, we discuss the strength 
required for the neutron-star magnetic field, and show that it needs to 
be moderately strong but not at a magnetar level. In Section 3, we 
discuss the applicability of scenarios (i)-(iii), finding that (i) is 
excluded but that (ii) and (iii) could be viable possibilities. In 
Section 4, we discuss the role of the propeller effect and transient 
behaviour, and Section 5 contains conclusions.

%%%%%%%%%%%%%%%%%%%%%%%%%%%%%%%%%%%%%%%%%%%%%%%%%%%%%%%%%%%%%%%%%%
%%%%%%%%%%%%%%%%%%%%%%%%%%%%%%%%%%%%%%%%%%%%%%%%%%%%%%%%%%%%%%%%%%

\section{Classical accretion estimates for the neutron-star 
magnetic field strength}

In the standard picture for HMXBs, the system has to be fairly young 
because the companion donor star is massive enough to have only a rather 
short main-sequence life-time. Matter coming from the donor star falls 
towards its neutron star companion, becomes included in a Keplerian 
accretion disc, and eventually becomes entrained by the neutron star's 
magnetic field, creating hot X-ray emitting accretion columns above the 
magnetic poles \citep[cf.][]{PR72}. Sufficiently young neutron stars 
typically have magnetic-field strengths above $10^{12} \, \rm{G}$ 
\citep[see, for example, the data in the ATNF pulsar catalogue,][]{ATNF}, 
with a tail of the distribution extending beyond $10^{13} \, \rm{G}$ and 
eventually joining with the magnetar regime at $10^{14} \, \rm{G}$. This 
can be relevant for explaining how this source can be so luminous, 
because if the magnetic field is stronger than the quantum limit, $B > 
4.4 \times 10^{13} \rm{G}$, the scattering cross-section would be 
suppressed, reducing the opacity of matter in the accretion columns above 
the magnetic poles and allowing higher luminosities. With this in mind, 
\citet{E14} suggested that this source might contain a magnetar. The 
computational results of \citet{M15} and the evidence of propeller effect 
from \citet{T15} support this idea. In other works, however, \citet{C14}, 
\cite{L14} and \cite{DPS14} have explored different scenarios with 
standard pulsar fields $\sim 10^{12} - 10^{13} \rm{G}$, while other 
authors have advocated weaker fields, $\sim 10^{9} \rm{G}$ \citep{KL14, 
T14}. As the present work was being completed, we have seen a new paper 
by \citet{KL16}, advocating a model with strong beaming and a magnetic 
field of $\sim 10^{11} \rm{G}$. The work presented here represents a line 
of study parallel to theirs.

We focus here on a scenario with a field at the top end of the range for 
standard pulsars. In the rest of this section, we apply some simple 
assumptions for testing the relevance of a solution of this type.

We take the full entrainment of the accreting matter by the magnetic 
field to occur close to the magnetic radius, $r_{\rm{m}}$, where the 
magnetic pressure balances the ram pressure of the infalling matter. 
Using the condition of mass continuity, we then obtain the following 
expression for the magnetic radius:
\begin{equation}
r_{\rm{m}}^{7} = \frac{B_{\rm{NS}}^4 R_{\rm{NS}}^{12}}{8 \zeta^2 G 
M_{\rm{NS}} 
\dot{M}^2} .
\label{eq:rm7}
\end{equation}
 Here $B_{\rm{NS}}$ is the field strength at the surface of the neutron 
star, and $\zeta$ is the ratio of the accretion velocity to the free-fall 
velocity \citep{WK89}. At $r = r_{\rm{m}}$, the accreting matter is taken 
to come into corotation with the neutron star, with the corotation speed 
being $v_{\rm{corot}} = 2 \pi r_{\rm{m}} P^{-1}_{\rm{s}} $.
 Since the system is probably close to spin equilibrium \citep{B14, 
DPS14}, it is reasonable to take $v_{\rm{corot}}$ as being approximately 
equal to the Keplerian velocity at $r = r_{\rm{m}}$. We can then estimate 
the appropriate value of the field strength in order to be consistent 
with the observed spin period for ULX M82 X-2. This gives
 \begin{eqnarray}
 B_{\rm{NS}} &=& 2^{-5/12} \pi^{-7/6} \zeta^{1/2} G^{5/6} 
M_{\rm{NS}}^{5/6} 
R_{\rm{NS}}^{-3} \dot{M}^{1/2} P_{\rm{s}}^{7/6} 
\nonumber \\
&=& 4.46 \times 10^{13}  \rm{G} 
\nonumber \\
&& \ \ \times \ \zeta^{1/2} \left( \frac{M_{\rm{NS}}}{1.4 M_{\odot}} 
	\right)^{5/6} 
	\left( \frac{R_{\rm{NS}}}{10^{6} \rm{cm}} \right)^{-3} 
\nonumber \\
&& \ \ \times 
	\left( \frac{\dot{M}}{2.0 \times 10^{20} \rm{g \, s}^{-1}} 
	\right)^{1/2} 
	\left( \frac{P_{\rm{s}}}{1.37 \rm{s}} \right)^{7/6} .
\label{eq:BNS}
\end{eqnarray}
Interestingly, this value is just above the critical quantum limit 
$B_{\rm{crit}}$ at which electron scattering is suppressed, as mentioned 
earlier:
 \begin{equation}
B_{\rm{crit}} = \frac{m_{\rm{e}}^2 c^3}{\hbar e} = 4.4 \times 10^{13} 
\rm{G} 
\end{equation}
\citep{E14}.

On the other hand, we can also consider the spin-up rate, focusing on the 
measured underlying tendency $\dot{P}_{\rm{s}} = -2 \times 10^{10} \rm{s 
\, s}^{-1}$ \citep{B14}, rather on the variations seen in the individual 
measurements, as mentioned earlier. We use the classical Ghosh \& Lamb 
model \citep{GL79}, in order to identify appropriate parameter values; 
the spin-up rate is then given by Eq. (15) in their paper. Assuming that 
the moment of inertia of the neutron star has the canonical value, 
$10^{45} \rm{g \, cm}^{2}$, and using the spin-up rate given by 
\citet{B14}, we find that the magnetic field strength corresponding to 
$\dot{P}_{\rm{s}} = -2 \times 10^{-10} \rm{s \, s}^{-1}$ would be
$
 B_{\rm{NS}} \left( \dot{P}_{\rm{s}} = -2 \times 10^{-10} \right) = 4 
\times 10^{13} \rm{G} 
$,
 which is very similar to the value obtained above, in 
Eq.~(\ref{eq:BNS}), from considerations of the spin period.

Based on these two independent estimates, we conclude that having a field 
of $\sim 4 \times 10^{13} \rm{G}$ can be consistent with the 
interpretation that M82 X-2 could be an example (albeit a rather extreme 
one) of previously-known classes of HMXBs (see, for example, \cite{K14}). 
A field of this strength can be just large enough to permit the quantum 
suppression of electron scattering as invoked by \citet{E14}.

%%%%%%%%%%%%%%%%%%%%%%%%%%%%%%%%%%%%%%%%%%%%%%%%%%%%%%%%%%%%%%%%%%%%
%%%%%%%%%%%%%%%%%%%%%%%%%%%%%%%%%%%%%%%%%%%%%%%%%%%%%%%%%%%%%%%%%%%%

\section{The accretion mode of ULX M82 X-2}

In this section, we examine the different possibilities mentioned in the 
Introduction for the accretion mode of ULX M82 X-2. The two mechanisms 
operating for standard HMXBs are: (i) accretion from a spherical wind 
emitted by the donor star, and (ii) accretion from a disc-shaped wind, 
thought to come from a ``decretion disc'' around the donor star (in a 
decretion disc, angular momentum is continually added to the inner edge 
of the disc from the rapidly rotating central star, and the matter drifts 
outwards rather than inwards). Type (i) occurs for OB-type HMXBs, while 
type (ii) fuels Be-type HMXBs. Additionally, although there is no 
observational evidence for this in standard HMXB studies, it is possible 
that Roche lobe overflow (RLOF) could be relevant here; we are referring 
to this as type (iii). \citet{B14} and \citet{E14} both suggested RLOF as 
a likely accretion mode for M82 X-2, but did not discuss it in detail. In 
a recent paper, however, \citet{SL15} presented an argument suggesting 
that a certain proportion of ULXs should indeed be accreting neutron 
stars with RLOF mass-transfer. Here, we will consider each of the 
mechanisms in turn to assess which of them may be appropriate for our 
case. In order to give a luminosity as high as that observed, the 
accretion rate needs to be at least as high as $\dot{M} \sim 10^{-6} 
M_{\odot} \rm{yr}^{-1}$, as shown earlier.

\subsection{Accretion fed from a spherical wind}

First, we consider whether a spherical stellar wind coming from the 
massive donor could successfully fuel ULX M82 X-2 (type (i) accretion). 
In simple stellar wind models, the wind velocity is frequently described 
by an expression of the form
\begin{equation}
v_{\rm{w}} \left( r \right) = v_{\infty} \left( 1 - \frac{R_{*}}{r} 
\right)^{\beta} 
\end{equation}
 \citep{KP00}, where $v_{\infty}$ is the terminal velocity of the wind, 
$R_{*}$ is the stellar radius, $r$ is the distance from the centre of the 
donor star, and $\beta$ is a parameter whose value is close to unity (we 
will take $\beta = 1$ exactly here). The accretion rate from a spherical 
wind is usually described by the Hoyle-Lyttleton formula \citep{HL39}:
\begin{equation}
\dot{M} = \rho_{\rm{w}} v_{\rm{rel}} \pi R_{\rm{acc}}^2 ,
\label{eq:dotM}
\end{equation}
where $R_{\rm{acc}}$ is the accretion radius defined by
\begin{equation}
R_{\rm{acc}} = \frac{2 G M_{\rm{NS}}}{v_{\rm{rel}}^2} ,
\label{eq:Racc}
\end{equation}
 and $v_{\rm{rel}}$ is the velocity of the neutron star relative to the 
accreting matter: $v_{\rm{rel}}^2 = v_{\rm{w}}^2 + v_{\rm{orb}}^2$, where 
$v_{\rm{orb}}$ is the orbital velocity of the neutron star. The wind 
density can be written as
\begin{equation}
\rho_{\rm{w}} \left( r \right) = \frac{\dot{m}_{\rm{w}}}{4 \pi r^{2} 
v_{\rm{w}}} , \label{eq:rhow}
\end{equation}
where $\dot{m}_{\rm{w}}$ is the rate of mass loss from the donor star. 
If one specifies a particular value for $\dot{m}_{\rm{w}}$, the accretion 
rate $\dot{M}$ can be calculated from Eq.~(\ref{eq:dotM}), and then the 
corresponding X-ray luminosity of the neutron star can be obtained from 
$L_{\rm{X}} = G M_{\rm{NS}} \dot{M} R_{\rm{NS}}^{-1}$.

\begin{figure}
\includegraphics[width=8cm]{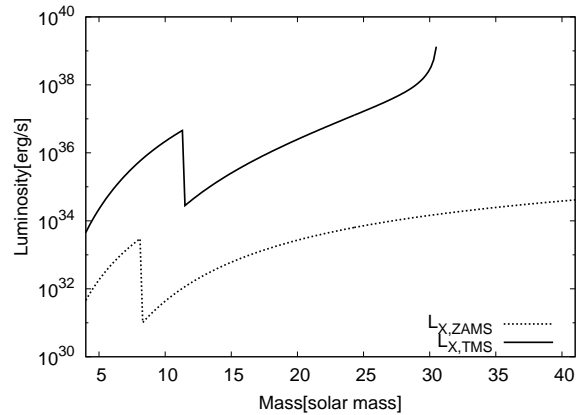}
\caption{
 Predicted X-ray luminosities are shown as functions of the mass of the 
donor star, assuming spherical wind accretion. The different curves 
correspond to different evolutionary phases: the zero-age main sequence 
phase (ZAMS, dashed curve) and the terminal main sequence phase (TMS, 
solid curve).
 } 
\label{fig:wind}
\end{figure}

\citet{V01} have given analytic formulae from which $\dot{m}_{\rm{w}}$ 
can be calculated for high-mass stars in terms of their mass $M_{\ast}$, 
luminosity $L_{\ast}$, effective temperature $T_{\rm{eff}}$ and the ratio 
$v_{\infty} / v_{\rm{esc}}$, where $v_{\infty}$ is the wind's terminal 
velocity and $v_{\rm{esc}}$ is the escape velocity from the stellar 
surface. A change in behaviour occurs when $T_{\rm{eff}}$ passes a value 
$T_{\rm{crit}} \approx 25000 \rm{K}$ and separate formulae are given for 
temperatures above and below that. 
Here we use the mass 
loss rates given by equations (24) and (25) in their paper.
The ratio $v_{\infty} / v_{\rm{esc}}$ is known to be $\sim 2.6$ for 
$T_{\rm{eff}} > T_{\rm{crit}}$ and $\sim 1.3$ for $T_{\rm{eff}} < 
T_{\rm{crit}}$ \citep{L95,V01} while values for $L_{\ast}$ and 
$T_{\rm{eff}}$ can be obtained as functions of $M_{\ast}$ using a stellar 
evolution code (we used the analytic fitting formulae from the paper by 
\citet{HPT00}). Putting this data into Eqs.~(\ref{eq:dotM}) -- 
(\ref{eq:rhow}), we could then calculate predicted values of the X-ray 
luminosity for a sequence of masses of the donor star, and our results 
are shown in Fig.~\ref{fig:wind}.

The two curves in Fig.~\ref{fig:wind} correspond to different 
evolutionary phases; the zero-age main sequence phase (solid curve) and 
the terminal main sequence phase (dashed curve). For stars more massive 
than 30 $M_{\odot}$, the stellar radius becomes larger than that of the 
binary orbit,
\begin{equation}
a = \left[ \frac{P_{\rm{orb}}^2}{4 \pi^2} G \left( M_1 + M_{\rm{NS}} \right) 
\right] ^{1/3} ,
\label{eq:a}
\end{equation}
 before it reaches the terminal main sequence (here $M_1$ is the mass of 
the donor star). It is clear, from this figure, that even if we consider 
the most efficient conditions, 
and an optimal viewing angle,
an X-ray luminosity of ULX level ($L_{\rm{X}} \simeq 10^{40}$) 
cannot be achieved. 
Hence, we conclude that ULX M87 X-2 cannot be fed by a spherical 
wind because the density of the wind would not be high enough.

\subsection{Accretion fed by a decretion-disc wind coming from around a 
Be star}

\begin{figure}
\includegraphics[width=8cm]{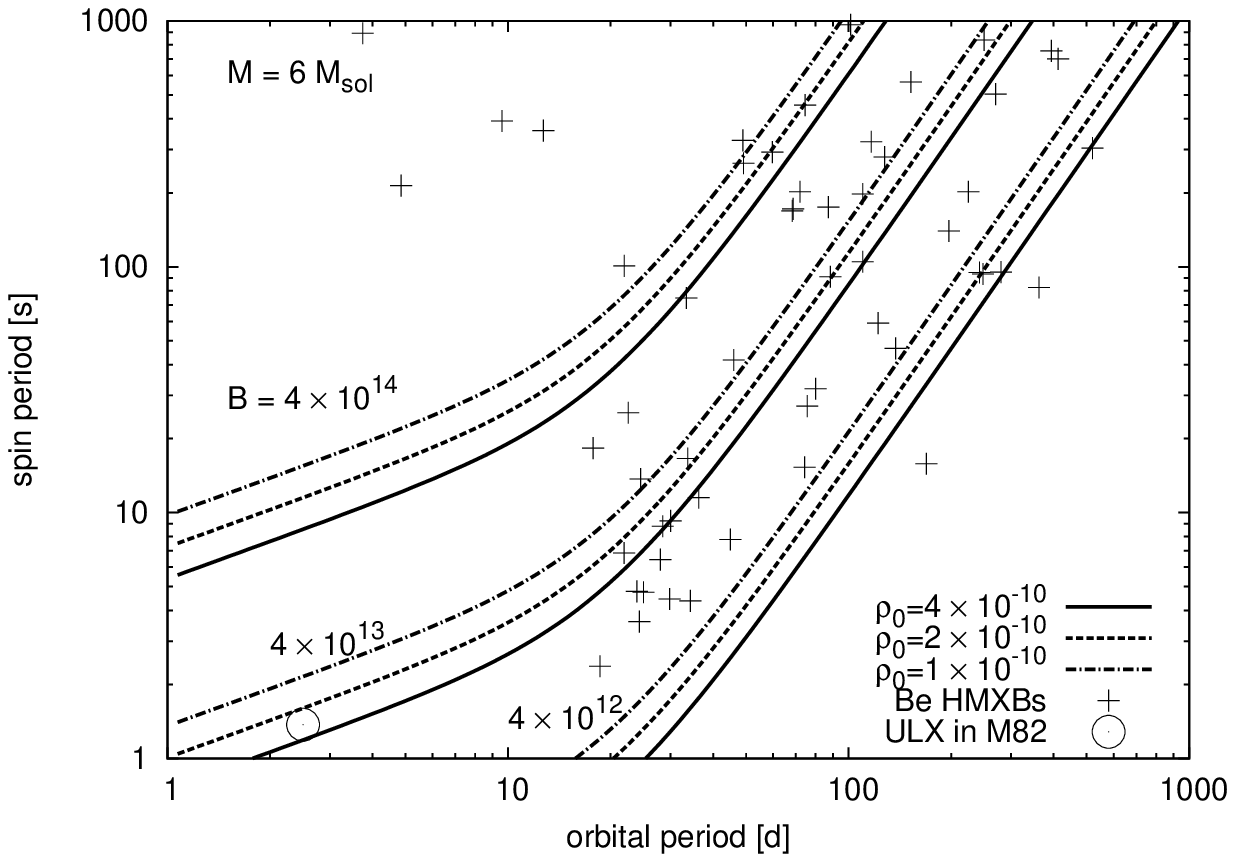}
\includegraphics[width=8cm]{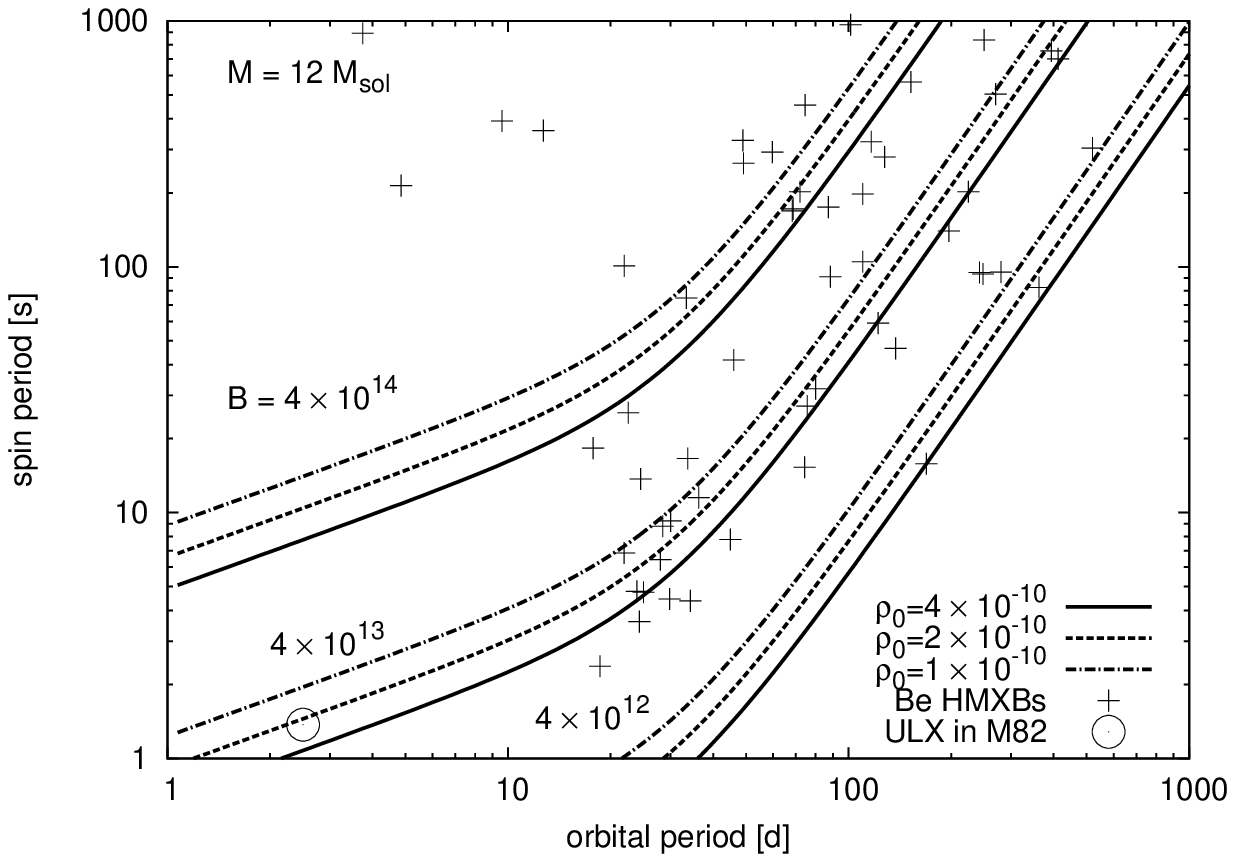}
\caption{
The Corbet diagram, showing the orbital period of the binary system 
$P_{\rm{orb}}$ plotted against the spin period of the neutron star 
$P_{\rm{s}}$. The upper and lower panels show results for donor masses of 
$6 M_{\odot}$ and $12 M_{\odot}$, respectively. The 
$P_{\rm{orb}}$-$P_{\rm{s}}$ relation given by equation (\ref{eq:PP}) is 
shown by the continuous curves with the three sets of curves in each 
panel showing results for different values of $B$ (measured in gauss), 
and the different curves of each set being for different disc densities 
(in $\rm{g \, cm}^{-3}$). The stellar radii are set to the terminal main 
sequence values obtained from the stellar evolution tracks of 
\citet{HPT00}. The positions on the diagram of known standard Be-type 
HMXBs are shown with crosses, while that for ULX M82 X-2 is shown with a 
circle. The data for the Be-HMXBs is taken from \citet{K07} and 
\citet{K14}.
 } 
\label{fig:Be}
\end{figure}

The Corbet diagram (where $P_{\rm{orb}}$ is plotted against $P_{\rm{s}}$) 
is frequently used in studies of HMXBs, and it can be useful to consider 
it also in the present context. OB-type and Be-type HMXBs show clearly 
different distributions in these diagrams, with the majority of the 
Be-type ones being close to the diagonal from lower-left to upper-right 
(they are the points marked with a $+$ in Fig.~\ref{fig:Be}). 
\citet{WK89} obtained an expression for the sequence followed by the 
Be-type HMXBs by assuming the Hoyle-Lyttleton accretion rate given by 
Eq.~(\ref{eq:dotM}). The decretion disc wind parameters $\rho_{\rm{w}}$ 
and $v_{\rm{w}}$ could be taken from any appropriate wind model; 
\citet{WK89} adopted a simple disc-shaped model:
\begin{eqnarray}
\rho (r) = \rho_0 \left( \frac{r}{R_{\rm{Be}}} \right)^{-n} , \\
v (r) = v_0 \left( \frac{r}{R_{\rm{Be}}} \right)^{n-2} ,
\end{eqnarray}
 where $\rho_0$ and $v_0$ are the density and velocity at the stellar 
radius $R_{\rm{Be}}$, and $n$ is a constant which needs to be fixed. In 
order to explain the positions of known standard Be-HMXBs in the Corbet 
diagram, \citet{WK89} used $n = 3.25$, and we will also use this value in 
the following. This choice is roughly consistent with recent numerical 
computations of Be-disc winds \citep{K11, C12}. With this wind model and 
any specified value of the magnetic field strength, one can obtain a 
corresponding relationship between $P_{\rm{s}}$ and $P_{\rm{orb}}$ by 
substituting equation (\ref{eq:dotM}) into equation (\ref{eq:BNS})  
($P_{\rm{orb}}$ is related to $a$ via equation (\ref{eq:a})). This gives
\begin{equation}
P_{\rm{s}} = C \left( \frac{v_{\rm{w}}}{v_{\rm{rel}}} \right)^{-9/7}
	\left( \frac{\pi \rho_0}{v_0^3} \right)^{-3/7} 
        \left( \frac{a}{R_{\rm{Be}}} \right)^{3(4n-6)/7} ,
\label{eq:PP}
\end{equation}
where
\begin{equation}
C = 2^{15/14} \pi \left( B_0 R_{\rm{NS}}^{3} \right)^{6/7} \left( 2 G 
M_{\rm{NS}} \right)^{-11/7} .
\end{equation}

In Fig.~\ref{fig:Be} we show $P_{\rm{orb}}$-$P_{\rm{s}}$ curves, obtained 
from Eq.~(\ref{eq:PP}), for three values of both $B$ and $\rho_0$. The 
values used for $v_0$ and the radius of the neutron star are the same as 
those used by \citet{WK89}: $v_0 = 10^6 \rm{cm \, s}^{-1}$ and 
$R_{\rm{NS}} = 10^{6} \rm{cm}$. In the two frames, we show the results 
for two different donor masses: $6 M_{\odot}$ and $12 M_{\odot}$, with 
the corresponding radii being those for the terminal main sequence phase 
on the stellar evolution tracks of \citet{HPT00}: $R_{\rm{Be}} = 10.7 
R_{\odot}$ for the $12 M_{\odot}$ donor and $R_{\rm{Be}} = 6.8 R_{\odot}$ 
for the $6 M_{\odot}$ donor. It can be seen that the results in the two 
frames are very similar. In each of them, ULX M82 X-2 (marked with the 
circles) takes a rather extreme position in comparison with the Be-HMXBs 
but, for an appropriate set of parameters, it does fit with the Be-HMXB 
sequences. As shown in the figure, assuming a magnetic field strength of 
$B = 4 \times 10^{13} \rm{G}$, the model given by Eq.~(\ref{eq:PP}) with 
the density $\rho_{0} = 2 - 3 \times 10^{-10} \rm{g \, cm}^{-3}$ does fit 
the position of ULX M82 X-2 very well in both cases.

Of course, the position of ULX M82 X-2 on the figure could also be 
explained by models with lower $B$ and extremely low $\rho_0$, or higher 
$B$ and extremely high $\rho_0$. From the limited available polarization 
data, \citet{D14} suggested that the $\rho_0$ for discs around Be-stars 
is between $8 \times 10^{-11}$ and $4 \times 10^{-12} \rm{g \, cm}^{-3}$, 
based on the nine systems which they studied. \citet{T11} tested values 
of $\rho_0$ between $2 \times 10^{-12}$ and $2 \times 10^{-10} \rm{g \, 
cm}^{-3}$ for fitting the IR radiation from Be discs, and suggested that 
the best fit value is $1.4 \times 10^{-10} \rm{g \, cm}^{-3}$. In view of 
these results, our best fit density ($\rho_{0} = 2 \times 10^{-10} \rm{g 
\, cm}^{-3}$) seems slightly high but still a reasonable value. 
Additionally, if we take this value for $\rho_0$, Eq.~(\ref{eq:dotM}) 
gives $3.2 \times 10^{20} \rm{g \, s}^{-1}$ for the mass accretion rate 
which is again quite reasonable for ULX M82 X-2. We note that a field as 
low as $\sim 10^{11} \rm{G}$ would be difficult to accommodate within 
this picture.

With this large mass loss rate, however, the donor Be-star cannot 
maintain a high density disc for very long and so, after the bright (high 
accretion rate) phase, it would run out of decretion disc matter. Hence, 
in this scenario, the accreting system would inevitably become a 
transient. Since, according to the archival survey, ULX M82 X-2 does show 
transient tendencies, this prediction seems to be consistent with the 
observations \citep{DSD14}. We note, however, that transient behaviour 
could also be understood in the context of propeller switching 
\citep{T15}, as we will discuss later.

\subsection{Roche lobe overflow}

\citet{B14} suggested that the accretion mode for ULX M82 X-2 may involve 
Roche lobe overflow (RLOF) and we need to consider this scenario as well.

\begin{figure}
\includegraphics[width=8cm]{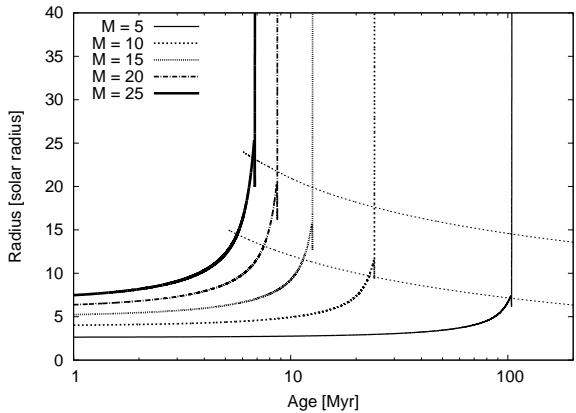}
\caption{
The radii of potential donor stars, with selected masses, are plotted as 
a function of their age. Each track starts from the zero-age main 
sequence and terminates at the starting point of the giant branch. The 
dotted curves link the points where each track crosses its corresponding 
Roche radius (lower curve) and orbital radius (upper curve). Data from 
\citet{HPT00}. 
} 
\label{fig:radii}
\end{figure}

\begin{figure}
\includegraphics[width=8cm]{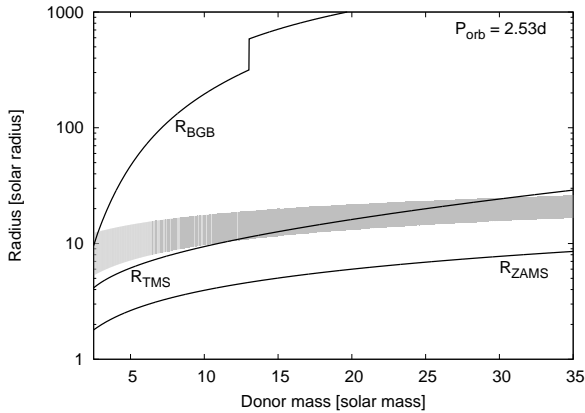}
\caption{
 Relevant radii are plotted as functions of the mass of the donor star. 
The Roche lobe radius and orbital radius are shown bounding the shaded 
region below and above, respectively. The radii of potential donor stars 
are shown at various evolutionary stages: on the zero-age main sequence 
(lower curve), the end of the main sequence (middle curve) and the 
beginning of the giant branch (upper curve). The data are taken from 
equations (1) -- (30) of \citet{HPT00}. 
} 
\label{fig:RL}
\end{figure}

An approximate value for the Roche radius $R_{\rm{RL}}$ is given by 
\begin{equation}
R_{\rm{RL}} = \frac{0.49 q^{2/3}}{0.6 q^{2/3} + \ln \left(1+ q^{1/3} 
\right) } a
\end{equation}
 \citep{E83}, where $q$ is the mass ratio $M_1 / M_{\rm{NS}}$ (with $M_1$ 
being the mass of the donor star) and $a$ is an orbital separation given 
by Eq.~(\ref{eq:a}). For having RLOF accretion, the donor radius needs to 
be larger than the Roche radius but smaller than the orbital radius. 
Fig.~\ref{fig:radii} shows evolutionary tracks for the radii of stars 
with selected masses (again using data from \citet{HPT00}) with the 
dotted lines linking the points where each track crosses its 
corresponding Roche radius (lower curve) and orbital radius (upper 
curve). The shaded region in Fig.~\ref{fig:RL} shows where RLOF can occur 
and is plotted as a function of the mass of the donor star. The upper 
bound of this shaded region corresponds to the orbital radius, and the 
lower bound shows the Roche radius. Stellar radii at several evolutionary 
stages (zero-age main sequence, the end of the main sequence and the 
beginning of the giant branch) are also shown.

If $M_1$ is larger than $11 M_{\odot}$, the donor star already fills its 
Roche lobe during the main sequence stage, leading to the type of mass 
transfer known as Case A \citep{P71}. This occurs on the thermal 
timescale
\begin{equation} 
\tau_{\rm{th}} \sim \frac{GM^2}{RL} . 
\label{eq:tauth} 
\end{equation} 
For a massive star, this timescale is shorter than $10^6 \rm{yr}$ and 
the mass transfer rate reaches $10^{-5} M_{\odot} \rm{yr}^{-1}$, which is 
sufficient for feeding ULX M87 X-2. However, for the most massive stars 
($M \ge 30 M_{\odot}$), the main-sequence stellar radius would be larger 
than the orbital radius as well, so that a common envelope would form 
around the binary in a short time scale. In that case, the system would 
be luminous at infra-red frequencies rather than in X-rays \citep{I13}; 
the donor must therefore be less massive than this in order to give the 
right sort of RLOF accretion. Furthermore, the mass of the donor ought to 
be restricted by the fact that the companion is a neutron star: the 
progenitor of the neutron star would have been the primary when this 
binary system was born and, according to \citet{F99}, it should have had 
an original mass of less than $20 - 30 M_{\odot}$ in order to give rise 
to a neutron star rather than a black hole. The present primary should be 
less massive than the original one (even if its mass may have increased 
slightly during the first mass transfer stage), and hence the donor mass 
is constrained to be less than roughly $\sim 25 M_{\odot}$.

Another issue which needs to be taken into account is that if the mass 
ratio $q = M_{1} / M_{\rm{NS}}$ is above a certain limit ($\sim 5$ for a 
red giant, and $\sim 12$ for a high-mass main sequence star), the binary 
would be subject to the Darwin instability, meaning that it could not 
sustain a circular orbit as required in order for RLOF to be a viable 
candidate mechanism here \citep{C73, E06}.

For $M_1$ between $\sim 5 - 11 M_{\odot}$, the star will be inside its 
Roche lobe during its main sequence lifetime but may overflow it during 
its expansion at the end of the main sequence (Hertzsprung gap phase), 
giving an early Case B type of mass transfer. This also proceeds on the 
thermal timescale, as given above, and could in principle be sufficient 
to feed ULX M87 X-2. However, in general the timescale on which the star 
crosses the Hertzsprung gap is quite short. For instance, an $8 
M_{\odot}$ star evolves from the end of the main sequence to the giant 
branch in 0.35 Myr \citep{E06} and would have a radius between 
$R_{\rm{RL}}$ and $a$ for only some fraction of that (determining which 
would require a detailed stellar evolution calculation). The relevant 
time is then very short compared with the overall life-time of the star 
($\sim 40 \rm{Myr}$ for $8 M_{\odot}$). However, the duration of the 
Hertzsprung gap phase depends sensitively on the mass of the donor star 
and for the minimum donor mass envisaged, $5.2 M_{\odot}$ \citep{B14}, it 
becomes $\sim 1.5 \rm{Myr}$. While it would still require a lucky chance 
to see any ULX fed by this Case B type of RLOF accretion, it would become 
less unlikely with the least massive possible donor stars. In connection 
with this: according to the population synthesis simulation of 
\citet{SL15}, the number of ULXs containing a neutron star should peak 
for systems having donor stars in a small mass range around $6-8 
M_{\odot}$, and so it is possible that the short ULX lifetime of these 
systems could be compensated by there being so many of them, and that 
they might actually predominate in the observations.

To summarise: accretion via RLOF might possibly be the mechanism for ULX 
M82 X-2 if the donor is an evolved star of $\sim 5 - 8\, M_{\odot}$ 
passing through the Hertzsprung gap. An RLOF scenario with a main 
sequence donor of $\sim 12 - 20\, M_{\odot}$ could also be a possibility 
although, for the higher masses, such a system would be prone to the 
orbit becoming eccentric due to the Darwin instability. An RLOF scenario 
with the donor being evolved up to the giant phase is ruled out 
completely.

%%%%%%%%%%%%%%%%%%%%%%%%%%%%%%%%%%%%%%%%%%%%%%%%%%%%%%%%%%%%%%%%%%%%
%%%%%%%%%%%%%%%%%%%%%%%%%%%%%%%%%%%%%%%%%%%%%%%%%%%%%%%%%%%%%%%%%%%%

\section{Some Further Issues}

In the previous sections, we have discussed similarities between ULX M82 
X-2 and known standard HMXBs, in terms of the strength of the neutron 
star's magnetic field and the accretion mode. In Section 2, we concluded 
that a fairly strong magnetic field, of around $4 \times 10^{13} \rm{G}$ 
is favoured for this object. While our considerations have been quite 
simple ones, this conclusion is consistent with more detailed analyses 
\citep{E14,DPS14}. In Section 3, we considered the possible feeding 
mechanisms, concluding that accretion modes similar to those of standard 
Be-HMXBs are possible here, and that some mechanisms involving Roche lobe 
overflow would also be possible. 
In the present section, we discuss two further related issues.

\subsection{Propeller effect and transient behaviour}

\begin{figure}
\includegraphics[width=8cm]{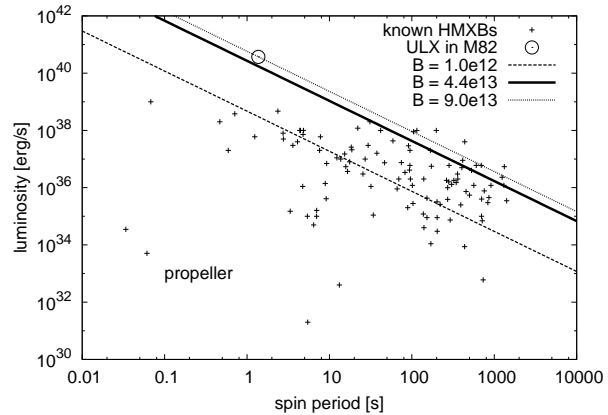}
\caption{
 The locations of ULX M82 X-2 and known HMXBs in the spin-period - 
luminosity plane. The critical lines of the propeller transition, for $B 
= 10^{12} \rm{G}$ (dashed), $4.4 \times 10^{13} \rm{G}$ (thick solid) and 
$9.0 \times 10^{13} \rm{G}$ (dotted) are shown. To avoid the propeller 
effect operating, the location of ULX M82 X-2 needs to be above the line 
corresponding the surface magnetic field strength of its neutron star, 
i.e. $B$ needs to be smaller than $9.0 \times 10^{13} \rm{G}$ for the 
parameter values given in the text, which is nevertheless significantly 
greater than the quantum limit $B \sim 4.4 \times 10^{13} \rm{G}$.
 } 
\label{fig:prop}
\end{figure}

The 1.37 second spin period of the neutron star in ULX M82 X-2 is shorter 
than those of most HMXBs (which are typically in the range of $\sim 
10-100 \rm{s}$). In view of this and the rather high magnetic field which 
we are inferring, one needs to consider whether it would be able to 
continue accreting matter from its disc, and continue to spin up, or 
whether it would enter the propeller regime where the disc matter is 
instead forced away \citep{IS75}. The critical period for onset of the 
propeller regime is given by \
\begin{equation} 
P_{\rm{s,crit}} = 81.5 \mu_{30}^{16/21} L_{\rm{X},36}^{-5/7} , 
\label{eq:PS} 
\end{equation} 
 where $\mu_{30}$ is the magnetic moment measured in units of $10^{30} 
\rm{G \, cm}^3$, as before, and $L_{\rm{X},36}$ is the X-ray luminosity 
in units of $10^{36} \rm{erg \, s}^{-1}$ \citep{I03,RP05}. 
Fig.~\ref{fig:prop} shows the spin period - luminosity plane, with the 
straight lines marking the relation given by Eq.~(\ref{eq:PS}) for 
various values of $B$. Also shown are the locations on the plot of known 
HMXBs (marked with crosses) and of ULX M82 X-2 (marked with the circle). 
For the latter, we are taking the values $P_{\rm{s}}=1.37 \rm{s}$ and 
$L_{\rm{X}}=3.7 \times 10^{40} \rm{erg \, s}^{-1}$. Below the line for 
the relevant value of $B$, the propeller regime is operative; above it, 
accretion can proceed. According to this, ULX M82 X-2 would avoid the 
propeller regime if it has $B < 9.0 \times 10^{13} \rm{G}$ and so field 
strengths between the quantum limit of $4.4 \times 10^{13} \rm{G}$ and 
the maximum of $9.0 \times 10^{13} \rm{G}$ could give accretion and 
spin-up, but magnetar-strength fields of above $10^{14} \rm{G}$ could not 
do so. The allowed range for $B$ to exceed the quantum critical limit but 
avoid the propeller effect, is small but non-vanishing.

It has been reported that archival data for M82 X-1/X-2 from XMM Newton 
shows only tiny fluctuations at a level below 2.2 \% \citep{DSD14}. This 
could be understood in two different ways. Firstly, the pulsations seen 
by \citet{B14} could be a transient behaviour so that while XMM Newton 
did not see the pulsations during the time-span of its observations, 
NuSTAR did see them. The second possibility is that the pulsations might 
be observable only in the high energy range above 10 keV, so that NuSTAR 
could detect them while XMM Newton could not. \citet{DPS14}, noting the 
variations seen by \citet{B14} in different measurements of 
$\dot{P}_{\rm{s}}$, have suggested that the system may be marginal for 
exhibiting the propeller effect, switching it on and off, and that the 
strong X-ray emissions and pulsations may only be observed when the 
propeller mechanism is not operating whereas during the propeller phase, 
only radiation from the hot disc is seen. 
In their latest and more extended analysis \citep{D16}, they conclude 
that the magnetically threaded disc model with $B \sim 10^{13} \rm{G}$ 
could explain the observed properties of this object  very well. 
\citet{T15} also argue that the 
observed bimodal distribution of X-ray luminosity can be interpreted as 
being due to the propeller effect switching on and off. In their picture, 
even when the system is in the propeller regime, a small fraction of 
matter still leaks through the magnetic field lines and continues to give 
rise to lower luminosity emission.

Another point is that HMXBs can often show different pulsation behaviour 
in different energy ranges, with the higher energy X-rays being emitted 
as a {\it{pencil beam}}, while lower energy ones are emitted as a 
{\it{fan beam}} \citep{BS75}. In this case, the emission directions would 
be different for the different energy ranges. In fact, more than 20\% of 
HMXBs show different light-curve shapes in high and low energy bands 
\citep{K07}. At present, the discussion on the transient behavior of M82 
X-2 cannot be concluded at all. More data is needed for deciding between 
these different possibilities.

\subsection{X-ray irradiation}

Also one needs to consider the effect of irradiation of the accreting 
matter by the strong X-ray emission coming from the neutron star. The 
direct effect of the radiation pressure coming from this is, of course, 
to act as another obstacle for the accretion. In order to achieve a high 
accretion rate in luminous X-ray systems, some means is required for 
avoiding this obstacle, such as anisotropy of the X-ray radiation 
\citep[see, for example,][]{C14,KL16}. In the wind accretion case, 
however, there are effects of the irradiation which act in the opposite 
direction, enhancing the accretion rate onto the neutron star. In 
detached wind-fed binaries, intra-binary matter cannot avoid some photo 
dissociation under these circumstances, and this reduces the efficiency 
of the acceleration mechanism driving the line-accelerated stellar wind 
away from the donor star \citep{SK90,W06,Ka14}. From equation 
(\ref{eq:Racc}) one can see that the resulting slower wind velocity would 
cause the accretion radius to be larger, and lead to \emph{increasing} 
the accretion rate onto the neutron star, as given by equation 
(\ref{eq:dotM}). However, the wind velocity is limited by $( 
v_{\rm{esc}}^2 + v_{\rm{orb}}^2 )^{1/2}$, where $v_{\rm{esc}}$ is the 
escape velocity at the donor surface, and so it cannot be slow enough to 
significantly affect our discussion here. Another potential positive 
effect of X-ray irradiation is that it could in principle cause the Roche 
lobe to become swamped with heated outer-envelope matter which could then 
overflow the Roche lobe even if the stellar radius was smaller than Roche 
radius \citep{ITF97}. This would require very strong irradiation though; 
it could only work in tight systems with orbital periods of a few hours, 
and would not play any significant role for M82 X-2.

%%%%%%%%%%%%%%%%%%%%%%%%%%%%%%%%%%%%%%%%%%%%%%%%%%%%%%%%%%%%%%%%%%%%
%%%%%%%%%%%%%%%%%%%%%%%%%%%%%%%%%%%%%%%%%%%%%%%%%%%%%%%%%%%%%%%%%%%%

\section{Conclusions}

In this paper, we have discussed the magnetic field strength and 
accretion mode for the recently identified neutron star in the source ULX 
M82 X-2.

We have considered the conditions required for producing the observed 
values of the key parameters: $P_{\rm{s}} = 1.37 \rm{s}$, $L_{\rm{X}} = 
3.7 \times 10^{40} \rm{erg \, s}^{-1}$, $P_{\rm{orb}} = 2.53 \rm{d}$ and 
$\dot{P}_{\rm{s}} = - 2 \times 10^{-10} \rm{s \, s}^{-1}$, and have 
argued that a consistent explanation can be given involving a moderately 
strong magnetic field at around the quantum limit $B \approx 4 \times 
10^{13} \rm{G}$. Having a field strength above the quantum limit is 
favourable for explaining the high observed luminosity, because of the 
reduced opacity in this case. However, we note that there have been a 
number of other suggestions regarding the magnetic field strength 
\citep{C14, DPS14, D16,E14, KL14, L14, T14}; further observations and 
discussions are required for resolving the issue.

We then went on to examine whether the standard accretion modes for HMXBs 
can be appropriate for this object. We concluded that spherical wind 
accretion, which drives OB-type HMXBs, cannot be the mechanism here but 
that an extension of the standard mechanism for Be-type HMXBs can provide 
a natural explanation. We have shown that if the neutron star has a 
moderately strong magnetic field ($B \sim 4 \times 10^{13} \rm{G}$) and 
there is wind accretion from a reasonably high-density ($\rho_0 = 2 - 3 
\times 10^{-10} \rm{g \, cm}^{-3}$) decretion disc around the 
Be-companion, then all of the main properties of ULX M82 X-2, including 
its position on the Corbet diagram, would be consistent with the 
relations followed by standard Be-HMXBs. Roche lobe overflow accretion is 
also a possibility. If the donor star has sufficiently high mass, it 
should still be on the main sequence and Case A RLOF accretion could then 
feed the ULX; if it has lower mass, it would need to be undergoing Case B 
RLOF while passing through the Hertzsprung gap during its 
post-main-sequence expansion.

\section*{Acknowledgments}

We gratefully acknowledge helpful discussions with Odele Straub and Hitoshi 
Yamaoka during the course of this work.

%\appendix

\end{document}